\documentclass[3p,number]{elsarticle}

\usepackage[usenames,dvipsnames]{color}
\usepackage{amsmath}
\usepackage{multirow}
\usepackage{graphicx}
\usepackage{hyperref}
\usepackage{xspace}
\usepackage{subfig}
\usepackage{cases}
\usepackage{cleveref}

\graphicspath{{../floats/eps/}{../floats/pdf/}{../floats/jpg/}{../floats/png/}}
\DeclareGraphicsExtensions{.pdf,.jpg,.png}

\definecolor{black}{RGB}{0, 0, 0}
\definecolor{gray}{RGB}{121, 121, 121}
\definecolor{blue}{RGB}{0, 84, 147}
\definecolor{green}{RGB}{146, 144, 0}
\definecolor{mocha}{RGB}{147, 82, 0}
\definecolor{asparagus}{RGB}{146, 144, 0}

\makeatletter
\def\ps@pprintTitle{%
 \let\@oddhead\@empty
 \let\@evenhead\@empty
 \def\@oddfoot{}%
 \let\@evenfoot\@oddfoot}
\makeatother

\begin{document}


\begin{frontmatter}

\title{Identifying top football players and springboard clubs from a football player collaboration and club transfer networks}

\author{Matic Tribu\v{s}on}
\ead{mt0932@student.uni-lj.si}

\author{Matev\v{z} Leni\v{c}}
\ead{ml9497@student.uni-lj.si}

\address{University of Ljubljana, Faculty of Computer and Information Science, Ve\v{c}na pot 113, SI-1000 Ljubljana, Slovenia}


\begin{abstract}
We consider all players and clubs in top twenty world football leagues in the last fifteen seasons. The purpose of this paper is to reveal top football players and identify springboard clubs. To do that, we construct two separate weighted networks. Player collaboration network consists of players, that are connected to each other if they ever played together at the same club. In directed club transfer network, clubs are connected if players were ever transferred from one club to another.
To get meaningful results, we perform different network analysis methods on our networks. Our approach based on PageRank reveals \textit{Christiano Ronaldo} as the top player. Using a variation of betweenness centrality, we identify \textit{Standard Liege} as the best springboard club.

\end{abstract}


\end{frontmatter}


\section{Introduction} 
Football is probably the most popular sport in the world with around 265 million active players \cite{BCount} around the globe and even more people enjoy watching it. Every year a lot of money is spent by football clubs in attempt to build a strong team by buying good players from their rivals.
Most of the data available on official football unions or tournament websites normally only addresses a specific match, tournament or season. In order to collect the data for top leagues in the last few seasons, we need to look elsewhere.
A very interesting website from this perspective is \texttt{www.transfermarkt.co.uk}. It contains all the major leagues including all the clubs, rankings, players, information about the players and also their estimated market value.\\
In an attempt to analyse players and the connections between them we construct a large network of professional football players from different clubs in different leagues. We are particularly interested in the influence of the teammates on a football player and if it is possible to identify the best players, based on knowing with whom they play now and where they played in the past. Using these analyses we could be able to find out which players are the best according to different metrics.
In a football player network, two players are connected to each other if they have ever played for the same club. Such network can be represented by a bipartite graph consisting of clubs and players. Every player is connected to all the clubs he has played for and through the nodes that represent clubs we are able to see which players played together for a specific club. For simpler analysis we separate these problems and project the bipartite graph to a network constructed only from nodes representing football players. Two nodes are connected to each other if they were ever teammates. This is an undirected network. 
Apart from the analysis of players, we also want to identify the best springboard clubs that are the players entry point into the best football clubs in the world. Because we do not include information about clubs in the first network, we construct a second network. The second network is a club transfer network. Clubs from the top twenty leagues represent nodes which are connected if any player was ever transferred from one club to another. The direction of the edge points from the club that sold the player to the club that bought the player.\\
Preliminary analysis on an unweighted undirected player collaboration network shows that weighted networks are needed in order to extract information about the best players. We expect very well known football players to come on top when analysing a weighted player collaboration network. In order to identify springboard clubs a weighted directed club transfer network has to be constructed. Weights of the edges are calculated using different equations that take into account multiple metrics. Using those networks we identify the top players in the world of football and the top springboard clubs.

\section{Related Work}
As it has been pointed out in \cite{pena2012network}, football data is becoming more easily available in the past years since FIFA  has made more data regarding different matches available on their website. Many authors took advantage of that and constructed different networks to perform network analysis and gather information from the networks.
In \cite{pena2012network} the authors used some interesting approaches to reveal key players of a certain team, performing analysis on a passing network of a specific team. They showed we are able to identify different kinds of strategies of a team such as focusing passes on a single player or evenly distributing passes between all players in the team. They performed several analyses on a team passing network using very well known network analysis methods such as PageRank, Betweenness centrality and Closeness centrality.\\
Player contribution to a team was also analysed in \cite{duch2010quantifying}. They used a variation of betweenness centrality of the player with regard to opponent's goal, which authors denoted as flow centrality.
We use similar network theory methods, but we adapt them to test different theories.
In \cite{cotta2013network} they dug a little deeper but followed the same idea. They only concentrated on one specific team and constructed more networks for the same match, introducing the time dimension.\\
Although there are various papers regarding football network analysis, the majority of football networks are only considering a certain match or tournament. In this paper we construct a much larger network consisting of thousands of players.
In other team sports, such as cricket, some authors have already tried to identify the best individuals among all the players that played over a certain time period.
A very interesting networks considering sportsmen throughout several decades were analysed in \cite{mukherjee2012identifying, radicchi2011best}. In \cite{radicchi2011best} the authors attempted to find out who is the best player of tennis in history of this sport. We try to construct a somewhat similar network but since football is very different from tennis the networks still differ a lot. Since the main difference is that football is a team sport, we can not just link players based on their matches. Here players are connected based on their affiliation to a club.

\section{Methods}
\subsection{Data Extraction and Network Construction}
In this paper we analyse a large set of football players throughout the past fifteen seasons. In order to collect this data we use the site \texttt{www.transfermarkt.co.uk}, which is becoming the leading portal when it comes to football players and information about them. Several scripts are used to extract relevant data for different clubs and players. Network is constructed from players out of 20 most valuable football leagues from year 2001 to 2016. The leagues and their values are presented in Table \ref{fig:leagueValue}. \\
Using the gathered data we constructed two separate networks, first one consisting of football players and the other consisting of football clubs. Football player network is a player collaboration network where players are connected if they ever played together at the same club. It is an undirected weighted network consisting of 36,214 nodes and 1,412,232 edges.
Other basic network properties are shown in Table \ref{fig:playersNetwork}.
Club network is a directed transfer network between all the clubs in the top twenty world football leagues. Nodes represent clubs and a club is connected to another club if a player was ever transferred from the first club to the second club. It is a directed weighted network consisting of 330 nodes and 12,841 edges.
Other basic network properties are shown in Table \ref{fig:clubsNetwork}.

\begin{table}[!h]
\centering
\begin{tabular}{llcl}
\multicolumn{1}{c}{\textbf{Property}}         &  & \textbf{Value} &      \\ \cline{1-1}
\multicolumn{1}{|l|}{Nodes}                   &  & \multicolumn{2}{c}{36,214}    \\ \cline{1-1}
\multicolumn{1}{|l|}{Edges}                   &  & \multicolumn{2}{c}{1,412,232} \\ \cline{1-1}
\multicolumn{1}{|l|}{Fraction of nodes in LCC} &  & \multicolumn{2}{c}{99.96\%}     \\ \cline{1-1}
\multicolumn{1}{|l|}{Average degree}          &  & \multicolumn{2}{c}{78.02}     \\ \cline{1-1}
\multicolumn{1}{|l|}{Average clustering}      &  & \multicolumn{2}{c}{0.67}      \\ \cline{1-1}
\end{tabular}
\caption{Player collaboration network properties}
\label{fig:playersNetwork}
\end{table}

\begin{table}[!h]
\centering
\begin{tabular}{llcl}
\multicolumn{1}{c}{\textbf{Property}}         &  & \multicolumn{2}{c}{\textbf{Value}} \\ \cline{1-1}
\multicolumn{1}{|l|}{Nodes}                   &  & \multicolumn{2}{c}{330}                    \\ \cline{1-1}
\multicolumn{1}{|l|}{Edges}                   &  & \multicolumn{2}{c}{12.841}                 \\ \cline{1-1}
\multicolumn{1}{|l|}{Fraction of nodes in LCC} &  & \multicolumn{2}{c}{99.90\%}                  \\ \cline{1-1}
\multicolumn{1}{|l|}{Average degree}          &  & \multicolumn{2}{c}{77.82}                  \\ \cline{1-1}
\multicolumn{1}{|l|}{Average distance}        &  & \multicolumn{2}{c}{2.03}                   \\ \cline{1-1}
\end{tabular}
\caption{Club transfer network properties}
\label{fig:clubsNetwork}
\end{table}

\begin{figure}[!h]
  \centering
  \includegraphics[width=0.9\textwidth]{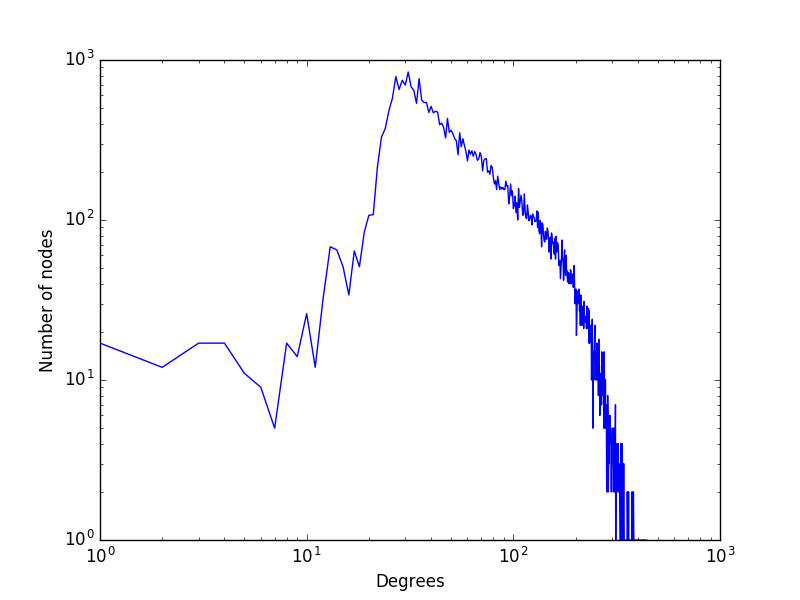}
  \label{fig:playersDistribution}
  \caption{Player network degree distribution}
\end{figure}

\begin{table}[!h]
\centering
\begin{tabular}{|c|c|l|c|c|}
\hline
\multicolumn{5}{|c|}{\textbf{Leagues}}                                                        \\ \hline
\textbf{League}      & \textbf{Value {[\pounds]}} &  & \textbf{League}     & \textbf{Value {[\pounds]}} \\ \hline
Permier League (ENG) & 3,01bn                 &  & Pro League (BEL)          & 354m             \\ \hline
La Liga (ESP)        & 2,25bn                 &  & Primera Division (ARG)    & 306m             \\ \hline
Serie A (ITA)        & 1,79bn                 &  & Premier Liga (UKR)        & 299m             \\ \hline
1. Bundesliga (GER)  & 1,65bn                 &  & Super League(GRE)         & 227m             \\ \hline
Ligue 1 (FRA)        & 1,06bn                 &  & Super League (SWI)        & 175m             \\ \hline
Super Lig (TUR)      & 698m                   &  & MLS (USA)                 & 162m             \\ \hline
Premier Liga (RUS)   & 638m                   &  & Liga 1 (ROM)              & 118m             \\ \hline
Serie A (BRA)        & 608m                   &  & 1. HNL (CRO)              & 115m             \\ \hline
Liga NOS (POR)       & 574m                   &  & Bundessliga (AUT)         & 110m             \\ \hline
Eredivisie (NED)     & 375m                   &  & Premiership (SCO)         & 85m              \\ \hline
\end{tabular}
\caption{Table showing value of top football leagues}
\label{fig:leagueValue}
\end{table}

\newpage

\subsection{Player Network Analysis}
In order to reveal the best players in our network, we choose an appropriate method of determining node importance. Since we wanted to identify the best players in the last fifteen seasons, we expected the most known and valued names of football to be at the top of the list. Not to neglect younger players, we also separate players into age groups. We analysed each age group individually in order to identify the most perspective players. Since our network is a collaboration network, we have to categorize the edges. The players that play with the best players are usually good themselves. Players with a lower value may change a lot of clubs and change a lot of teammates in a couple of seasons, but this categorization penalises their edges. In general, player market value is a good identifier of the quality of a player. Therefore we choose market value as a core property to calculate the edge weight. Since our data spans over fifteen seasons, we have to take into account the inflation, so that good players that played in the past are not penalised. We gather average inflation rate from \cite{InfRatio}. The final formula for calculating the weight of a specific edge is

\begin{equation} \label{eq:playersWeight}
weight = \sum\limits_{s}\frac{(pv_1 + pv_2) * (1 + (\theta \cdot (2016 - s)))}{100000} \quad .
\end{equation}

Symbols $pv_1$ and $pv_2$ are values of players that are connected by the edge, $s$ represents the seasons in which players played together and $\theta$ represents average inflation ratio per year for Europe in the last 13 years. The equation is divided by 100000, to obtain smaller numbers.
To calculate which node is the most important, we choose one of the most popular node importance algorithms, PageRank \cite{page1997pagerank}. We calculate the PageRank score of every node in our weighted network. To identify the most perspective players, we separate players into age groups. The most perspective players have the highest score in their age groups.

\subsection{Club Network Analysis}
From the club transfer network we want to identify the springboard clubs. These are the clubs where younger players gather experience and are later sold to better or even the best clubs in the world. Similar to the player collaboration network, this network has to be weighted as well. We are able to extract the number of transfers in both directions for all pairs of clubs but the absolute number does not provide the necessary information for springboard club identification. Thus, we have to weight every edge, representing the number of transfers from one club to another, with a weight related to the importance of the destination club. The importance of the destination club is calculated using two different equations. One is based on average ranking of the destination club in the past fifteen seasons and the ranking of the league they play in, and the other one is based on the destination club value. Both equations are stated and explained below.

\begin{equation} \label{eq:clubsWeightRanking}
weight = \frac{1}{r_c \times r_l}
\end{equation}
\begin{equation} \label{eq:clubsWeightValue}
weight = \frac{cv}{1000000}
\end{equation}

Weight in the Equation \ref{eq:clubsWeightRanking} is calculated as a reciprocal value of destination club average ranking in the past fifteen seasons $r_c$ multiplied by the our predefined destination club league ranking $r_l$. Predefined league rankings can be found in Table \ref{fig:leagueRankings} and are defined for the purpose of this paper.
Weight in the Equation \ref{eq:clubsWeightValue} is calculated as destination club average value in the past fifteen seasons $cv$ divided by $1000000$ to lower the weight values.\\
To identify springboard clubs we have to choose a different method from the one we use for player collaboration network. The most important thing in this network are the transfer paths from less valuable to the most valuable clubs. A club is considered a springboard if it is involved in a lot of transfers to the most valuable clubs. Thus, the betweenness centrality \cite{freeman1977set} is the most suitable measure. We implement a fast betweenness algorithm discussed in \cite{fastBet}. Since our network is weighted we have to modify the proposed algorithm so it takes weights into account. The only difference from the proposed algorithm is calculation of path lengths where we do not add one for every hop but take weight into account. We have to take the reciprocal value of weight as in our network larger weight is better and we want to favour edges with larger weights.

\begin{table}[!h]
\centering
\begin{tabular}{|c|c|l|c|c|}
\hline
\multicolumn{5}{|c|}{\textbf{League rankings}}          \\ \hline
\textbf{League}      & \textbf{Ranking} &  & \textbf{League}      & \textbf{Ranking} \\ \cline{1-2} \cline{4-5} 
La Liga (ESP)        & 100              &  & Premier League (UKR)      & 20               \\ \cline{1-2} \cline{4-5} 
Premier League (ENG) & 95               &  & Super League (SWI)        & 20               \\ \cline{1-2} \cline{4-5} 
Serie A (ITA)        & 85               &  & Serie A (BRA)             & 20               \\ \cline{1-2} \cline{4-5} 
Bundesliga (GER)     & 75               &  & Super Lig (TUR)           & 15               \\ \cline{1-2} \cline{4-5} 
Ligue 1 (FRA)        & 50               &  & Primera Division (ARG)    & 15               \\ \cline{1-2} \cline{4-5} 
Primera Liga (POR)   & 40               &  & Super League(GRE)         & 13               \\ \cline{1-2} \cline{4-5} 
Eredivisie (NED)     & 40               &  & Liga 1 (ROM)              & 12               \\ \cline{1-2} \cline{4-5} 
Pro League (BEL)     & 25               &  & 1. HNL (CRO)              & 10               \\ \cline{1-2} \cline{4-5} 
Premier League(RUS)  & 25               &  & Bundesliga (AUT)          & 10               \\ \cline{1-2} \cline{4-5} 
Premiership (SCO)    & 20               &  & MLS (USA) & 5                \\ \cline{1-2} \cline{4-5} 
\end{tabular}
\caption{Predefined league rankings (higher is better)}
\label{fig:leagueRankings}
\end{table}

\section{Results and Discussion}

\subsection{Top players}
After running the analysis on the player collaboration network, we can show that the best player according to our analysis is Cristiano Ronaldo. He is followed by several other players that have played for several of the best clubs. By looking at the Table \ref{fig:resultsPlayers}, where top 20 players identified by our algorithm and their scores are listed, we can see that the value of the player is not the only thing that affects the score of a player. Players like Beckham, Ronaldinho, Kak{\'a} and Keane, whose market value decreased a lot lately because of their age, but they played for a lot of important clubs in their career, have high scores. Most players on the top 20 list are still active today and are playing in the best leagues. \\
The most perspective players in each age group are listed in \Cref{fig:perspective9998,fig:perspective9796,fig:perspective9594,fig:perspective9392}.
When assessing player's perspectiveness, the most important factor besides his value and the values of his teammates is the player's age. Since our network is an undirected network connecting two players, age can not be simply added to the weight equation. Including age into weight equation would favour players that have valuable teammates and also players that have younger teammates, which is not desired. Therefore, for identifying the most perspective players, the network can stay the same, we just need to interpret results differently. We divide players into different groups based on their age and compare only scores of players in the same groups.
On average, older players have higher scores, which is expected as they played more seasons, which results in higher degree. Thus, the separation into age groups is beneficial. Some of the most perspective players based on our algorithm already play for the best clubs and others, despite their young age, play an important role in their clubs.\\
Based on the results, we can conclude that PageRank is an appropriate algorithm for determining the best players in our weighted network.

\begin{table}[!h]
\centering
\begin{tabular}{|c|c|c|}
\hline
\textbf{Player}    & \textbf{PageRank score} & \textbf{Value 2015/16 [\pounds]} \\ \hline
Cristiano Ronaldo      & 0.000557                & 77.000.000                   \\ \hline
Lionel Messi           & 0.000544                & 84.000.000                   \\ \hline
David Beckham          & 0.000528                & /             		        \\ \hline
Zlatan Ibrahimovi{\'c} & 0.000459                & 10.500.000                   \\ \hline
Ronaldinho Ga{\'u}cho  & 0.000444                & 1.005.000                    \\ \hline
Kak{\'a}               & 0.000417                & 3.500.000                    \\ \hline
Wayne Rooney           & 0.000407                & 28.000.000                   \\ \hline
Fernando Torres        & 0.000402                & 4.900.000                    \\ \hline
Steven Gerrard         & 0.000400                & 1.400.000                    \\ \hline
Samuel Eto'o           & 0.000399                & 1.400.000                    \\ \hline
Robbie Keane           & 0.000390                & 876.000                      \\ \hline
Daniele De Rossi       & 0.000389                & 5.250.000                    \\ \hline
Neymar                 & 0.000388                & 70.000.000                   \\ \hline
Cesc F{\'a}bregas      & 0.000377                & 35.000.000                   \\ \hline
Sergio Ag{\"u}ero      & 0.000376                & 42.000.000                   \\ \hline
Andr{\'e}s Iniesta     & 0.000376                & 24.500.000                   \\ \hline
Wesley Sneijder        & 0.000370                & 10.500.000                   \\ \hline
David Villa            & 0.000358                & 4.900.000                    \\ \hline
Gianluigi Buffon       & 0.000349                & 1.400.000                    \\ \hline
Carlos T{\'e}vez	   & 0.000347                & 14.000.000                   \\ \hline
\end{tabular}
\caption{Player collaboration network PageRank results}
\label{fig:resultsPlayers}
\end{table}

\begin{table}[!h]
\centering
\begin{tabular}{|c|c|c|c|}
\hline
\textbf{Player} & \textbf{PageRank score} & \textbf{Player}    & \textbf{PageRank score} \\ \hline
Gianluigi Donnarumma   & 0.000020       & Hachim Mastour       & 0.000023       \\ \hline
Alexandru Petrus       & 0.000011       & Ianis Hagi           & 0.000020       \\ \hline
Maximiliano Romero     & 0.000010       & Dani Olmo            & 0.000017       \\ \hline
Robert Moldoveanu      & 0.000009       & Martin {\"O}degaard  & 0.000015       \\ \hline
Vlad Dragomir          & 0.000009       & Reece Oxford         & 0.000015       \\ \hline
\end{tabular}
\caption{PageRank results for players born in year 1999 (left) and 1998 (right)}
\label{fig:perspective9998}
\end{table}

\begin{table}[!h]
\centering
\begin{tabular}{|c|c|c|c|}
\hline
\textbf{Player}    & \textbf{PageRank score} & \textbf{Player} & \textbf{PageRank score} \\ \hline
Youri Tielemans    & 0.000070       & Alen Halilovic   & 0.000052       \\ \hline
Breel Embolo       & 0.000054       & Gabriel          & 0.000052       \\ \hline
Malcom             & 0.000042       & Kingsley Coman   & 0.000050       \\ \hline
Ante \'{C}ori\'{c} & 0.000036       & Timo Werner      & 0.000049       \\ \hline
Andrija Bali\'{c}  & 0.000035       & Fabrice Olinga   & 0.000044       \\ \hline
\end{tabular}
\caption{PageRank results for players born in year 1997 (left) and 1996 (right)}
\label{fig:perspective9796}
\end{table}

\begin{table}[!h]
\centering
\begin{tabular}{|c|c|c|c|}
\hline
\textbf{Player}  & \textbf{PageRank score} & \textbf{Player} & \textbf{PageRank score} \\ \hline
Max Meyer        & 0.000058       & Mateo Kovacic    & 0.000107       \\ \hline
Luke Shaw        & 0.000057       & Marquinhos       & 0.000092       \\ \hline
Adrien Rabiot    & 0.000053       & Domenico Berardi & 0.000091       \\ \hline
{\'A}ngel Correa & 0.000052       & Raheem Sterling  & 0.000089       \\ \hline
Dorin Rotariu    & 0.000052       & Gerard Deulofeu  & 0.000086       \\ \hline
\end{tabular}
\caption{PageRank results for players born in year 1995 (left) and 1994 (right)}
\label{fig:perspective9594}
\end{table}

\begin{table}[!h]
\centering
\begin{tabular}{|c|c|c|c|}
\hline
\textbf{Player}    & \textbf{PageRank score} & \textbf{Player} & \textbf{PageRank score} \\ \hline
Romelu Lukaku      & 0.000202       & Neymar             & 0.000388       \\ \hline
Paul Pogba         & 0.000151       & Lucas              & 0.000178       \\ \hline
Julian Draxler     & 0.000143       & Mario G{\"o}tze    & 0.000175       \\ \hline
Rapha{\"e}l Varane & 0.000104       & Christian Eriksen  & 0.000157       \\ \hline
Luciano Vietto     & 0.000096       & Jack Wilshere      & 0.000148       \\ \hline
\end{tabular}
\caption{PageRank results for players born in year 1993 (left) and 1992 (right)}
\label{fig:perspective9392}
\end{table}

\newpage

\subsection{Springboard Clubs Identification}
From the club transfer network analysis we can show that the best springboard club among the clubs in the top twenty leagues is Standard Liege. The analysis provides very good results, since the top 15 clubs list is lacking the most valuable and the best clubs in the world. Top 15 clubs by betweenness centrality scores and their scores calculated on network using both weight equations are listed in Table \ref{fig:resultsClubs}. The results also show very slight difference between both proposed weight equations. The top two clubs are the same regardless of the weight and the third and the fourth switch positions if we change the weight calculation equation. All the clubs on the top 15 list are from less valuable leagues and these clubs normally buy younger players that are more affordable and sell the ones whose value rises above a certain level. This makes them a perfect springboard for younger and less experienced players. Because of such transfer activity such clubs get high score according to betweenness centrality as they play an important role in the transfer paths from less valuable clubs to the best clubs.

\begin{table}[!h]
\centering
\begin{tabular}{|c|c|l|c|c|}
\cline{1-5}
     \multicolumn{5}{|c|}{\textbf{Club ranking using betweenness centrality}}                            \\ \cline{1-5} 
     \textbf{Club} & \textbf{Score by value (Eq. \ref{eq:clubsWeightValue})} &  & \textbf{Club} & \textbf{Score by rank (Eq. \ref{eq:clubsWeightRanking})} \\ \cline{1-2} \cline{4-5} 
Standard Liege     & 0.013605                  &  & Standard Liege       & 0.012823               \\ \cline{1-2} \cline{4-5} 
AEK Athens         & 0.011217                  &  & AEK Athens           & 0.012240               \\ \cline{1-2} \cline{4-5} 
SL Benfica         & 0.010937                  &  & Sporting CP          & 0.010424               \\ \cline{1-2} \cline{4-5} 
Sporting CP        & 0.010312                  &  & SL Benfica           & 0.010172               \\ \cline{1-2} \cline{4-5} 
Skoda Xanthi       & 0.009605                  &  & AS Monaco            & 0.009275               \\ \cline{1-2} \cline{4-5} 
Dinamo Bukarest    & 0.008743                  &  & FC Porto             & 0.008988               \\ \cline{1-2} \cline{4-5} 
AS Monaco          & 0.008704                  &  & Rubin Kazan          & 0.008884               \\ \cline{1-2} \cline{4-5} 
Dinamo Zagreb      & 0.008675                  &  & CFR Cluj             & 0.008681               \\ \cline{1-2} \cline{4-5} 
Olympiacos Pir.    & 0.008553                  &  & Skoda Xanthi         & 0.008638               \\ \cline{1-2} \cline{4-5} 
CFR Cluj           & 0.008542                  &  & Dinamo Bukarest      & 0.008518               \\ \cline{1-2} \cline{4-5} 
Steaua Bucharest   & 0.008180                  &  & Olympiacos Pir.      & 0.008397               \\ \cline{1-2} \cline{4-5} 
Udinese Calcio     & 0.007899                  &  & Rangers FC           & 0.008216               \\ \cline{1-2} \cline{4-5} 
FC Porto           & 0.007889                  &  & Dinamo Zagreb        & 0.008170               \\ \cline{1-2} \cline{4-5} 
Celtic FC          & 0.007849                  &  & Iraklis Thess.       & 0.007925               \\ \cline{1-2} \cline{4-5} 
Petrolul Ploiesti  & 0.007794                  &  & Red Bull Salzburg    & 0.007907               \\ \cline{1-2} \cline{4-5} 
\end{tabular}
\caption{Club transfer network betweenness centrality results}
\label{fig:resultsClubs}
\end{table}

\newpage

\section{Conclusion}
Player collaboration network from the past fifteen seasons from the top twenty football leagues consists of over 36 thousand nodes and nearly 1.5 million edges. Therefore, time and space consuming algorithms can prove too demanding to run on regular computers. Weighted PageRank algorithm however was able to calculate the scores for all the players in a very reasonable time. With the PageRank algorithm and proper edge weight, we are able to identify the top players from the period of last fifteen seasons. A very important factor in the weight equation is the inflation rate which ensures that older players that were never as valuable as the best players of the last seasons are also present on the top players list.\\
Using the same network, we are also able to identify the most perspective football players by separating their PageRank scores into age groups. Using this approach, we compare only players of similar age that have played for similar number of seasons. This ensures the same conditions for all the players in a specific age group. Results highlight some young players that already play for the best football clubs and some young players from less known clubs, where they play an essential role.\\
Results from club transfer network analysis are very similar to initial hypothesis. We expect clubs from less valuable leagues to come on top. We are able to identify springboard clubs by using the data about player transfers from the past fifteen seasons by constructing a directed weighted network with adequate weights using the data we have on the club value or the club rankings in the past seasons. With the proposed network, we use a weighted betweenness centrality algorithm to reveal the best springboard clubs in the top football leagues in the world. Our algorithm identifies some clubs from Belgian, Greek and Portuguese leagues as the best springboard clubs.\\


\bibliographystyle{elsarticle-num}


\end{document}